\documentclass[11pt,aps,nofootinbib,showpacs,longbibliography,preprintnumbers,
notitlepage]{revtex4-1}

\usepackage{amssymb,graphicx,epsfig,amsmath,bm,natbib} 

\def\be{\begin{equation}}
\def\ee{\end{equation}}
\def\bea{\begin{eqnarray}}
\def\eea{\end{eqnarray}}
\begin{document}

\title{Do new data on ${\mathbf B^+} \to {\mathbf \tau^+} 
\mathbf{\nu_\tau}$ decays point to an early discovery of supersymmetry 
at the LHC?}

\author{Biplob Bhattacherjee}
\email{biplob,\,amol,\,diptimoyghosh,\,sreerup@theory.tifr.res.in} 
\affiliation{Department of Theoretical Physics, Tata Institute of 
Fundamental Research, 1, Homi Bhabha Road, Mumbai 400 005, India}

\author{Amol Dighe}
\email{biplob,\,amol,\,diptimoyghosh,\,sreerup@theory.tifr.res.in} 
\affiliation{Department of Theoretical Physics, Tata Institute of 
Fundamental Research, 1, Homi Bhabha Road, Mumbai 400 005, India}

\author{Diptimoy Ghosh}
\email{biplob,\,amol,\,diptimoyghosh,\,sreerup@theory.tifr.res.in} 
\affiliation{Department of Theoretical Physics, Tata Institute of 
Fundamental Research, 1, Homi Bhabha Road, Mumbai 400 005, India}

\author{Sreerup Raychaudhuri}
\email{biplob,\,amol,\,diptimoyghosh,\,sreerup@theory.tifr.res.in} 
\affiliation{Department of Theoretical Physics, Tata Institute of 
Fundamental Research, 1, Homi Bhabha Road, Mumbai 400 005, India}

\date{\today}

\begin{abstract}
The recent Belle and BaBar measurements of the branching ratio of $B^+ 
\to \tau^+ \nu_\tau$ indicate a significant deviation from the standard 
model prediction. We demonstrate that this measurement has a serious 
impact on models with minimal flavor violation involving a charged Higgs 
boson, ruling out a large portion of the currently allowed parameter 
space. In the constrained minimal supersymmetric standard model, this 
creates a tension between the measurements of $B^+ \to \tau^+ \nu_\tau$ 
and the anomalous magnetic moment of the muon, unless $\tan\beta$ is 
small, $\mu > 0$, and $A_0$ takes a large negative value. In fact, a very 
small region of the parameter space of this model, with small values of 
$m_0$ and $m_{1/2}$, survives all the constraints at 95\% C.L. It is 
remarkable that this specific region is still consistent with the 
lightest supersymmetric particle as the dark-matter. Moreover, it 
predicts observable supersymmetric signals in the early runs of the LHC, even 
perhaps at 7~TeV. We also show that a consistent explanation for the 
deviation of the $B^+ \to \tau^+ \nu_\tau$ branching ratio from the 
standard model can be achieved in a nonuniversal Higgs-mass model, 
which could also predict early signals of supersymmetry at the LHC.
\end{abstract}

\preprint{TIFR-TH/10-35}

\maketitle

\section{Introduction}

Now that the CERN Large Hadron Collider (LHC) has commenced its 
long-awaited run and the first physics results have been analyzed and 
made public \cite{LHC2010a,LHC2010b}, there is an atmosphere of palpable 
suspense in the high energy physics community as to what physics results 
will come out as more and more data are collected and studied, and most 
importantly, as to whether these results will indicate new physics (NP) 
beyond the Standard Model (SM). The experimental programme is more or 
less clear: more statistics will be accumulated, and the results will be 
compared with the predictions of the SM. Deviations from the latter 
would imply some sort of NP, and one can refer to existing theoretical 
studies to indicate what kind of NP is indicated by the observed 
deviation. It is true that theorists have not succeeded in providing an 
unequivocal prediction in this regard. This is because there exist 
several rival possibilities for NP, each with good arguments both for 
and against it. However, for several technical and aesthetic reasons, of 
which tractable ultraviolet behavior and the natural appearance of 
chiral fermions are perhaps the most important, supersymmetry (SUSY) has 
always been the pick of these NP models. At the dawn of the LHC era, it 
still remains the first option for any study of NP predictions.

Elegant as SUSY may be as an abstract idea, it is well known that it 
presents a very different face when it comes to constructing realistic 
models at low energies. Any phenomenologically viable SUSY model must 
necessarily include a large number of soft SUSY-breaking parameters. A 
count of the number of free phenomenological parameters in the so-called 
minimal supersymmetric standard model (MSSM) \cite{Nilles:1983ge, 
*Haber:1984rc, *Martin:1997ns, *Peskin:2008nw} runs to over 100, 
including masses, coupling constants, and mixing angles for the large 
number of supersymmetric partners, or sparticles, in the model. This 
proliferation of parameters may be directly traced to the fact that the 
MSSM does not include a specific mechanism for the breaking of SUSY, and 
hence, the numerous SUSY-breaking parameters are essentially put in by 
hand. Although such a model can exist, at least in principle, a theory 
with a hundred odd free parameters is a phenomenologist's nightmare, 
since it leads to very few clear predictions at the empirical level. At 
the LHC, for example, this leads to a wide landscape of possible signals 
which would leave an experimentalist with hard data to compare with a 
bewildering variety of options \cite{Konar:2010bi}. 
It is also difficult to believe that the 
breaking of SUSY is a sheer accident brought on by a proliferation of 
arbitrary nonzero parameters. One would rather argue that there is a 
definite mechanism for SUSY breaking \cite{wessbagger}, and when we know 
it, we will also know the parameters in question. Once again, however, 
theorists have failed to come up with an unambiguous mechanism for 
SUSY breaking -- there exist quite a few different suggestions 
\cite{godbole}, beginning with minimal supergravity (mSUGRA) models, 
through gauge-mediated SUSY breaking (GMSB), anomaly mediated 
SUSY breaking (AMSB) and so on, each with a very different pattern for 
the parameters in question. Each of these models has different 
predictions for LHC signals, and hence, in effect, the chaotic situation 
within the subset of SUSY models becomes a cameo of the general NP 
scenario.

The oldest, and perhaps the most restrictive, of these SUSY models where 
a specific mechanism for SUSY breaking is considered, is the so-called 
``constrained`` MSSM (cMSSM), which is based on an underlying mSUGRA \cite{Cremmer:1978iv, *Cremmer:1978hn, 
*Barbieri:1982eh, *Chamseddine:1982jx, *Hall:1983iz, *Nath:1983aw, 
*Ohta:1982wn} scenario\footnote{This mechanism, by invoking 
supergravity, gives up the good ultraviolet behavior, unless, indeed, 
the supergravity model is equivalent to, or embedded in, a string 
theory. Aficionados of the cMSSM would, of course, argue that gravity 
must eventually be included anyway.}. In this scenario, supergravity is 
broken spontaneously in a so-called ''hidden'' sector consisting of fields 
which do not have strong or electroweak couplings to the MSSM fields. 
However, gravity, which necessarily couples to all fields so long as 
they carry energy and momentum, acts as a mediator between the hidden 
sector and the MSSM sector, giving rise to the soft SUSY-breaking 
parameters. It is this circumstance that leads to a dramatic reduction 
in the number of parameters, since gravity is blind to all flavor and 
color quantum numbers, though it can sense the spin of a particle. As a 
result, the mSUGRA model has just five free parameters, viz. (i) a 
universal scalar mass $m_0$, (ii) a universal fermion mass $m_{1/2}$, 
(iii) a universal trilinear (scalar) coupling $A_0$, (iv) the ratio of 
vacuum expectation values of the two Higgs doublets, parametrized as 
$\tan\beta$, and (v) the Higgsino mixing parameter $\mu$. This 
universality of the masses and couplings is valid at the scale where the 
SUSY-breaking parameters are generated, which is usually identified with 
the scale of grand unification (GUT scale)\footnote{There exists a 
symbiotic relation between SUSY and grand unified theory (GUT) ideas, since SUSY solves the 
hierarchy problem in GUT , and a GUT is natural at the scale 
where SUSY breaking is generated.}, i.e., above $10^{16}$~GeV. While 
running down to low energies using the renormalization group (RG) 
equations, however, the various soft SUSY-breaking parameters evolve 
differently, and lead to a specific mass spectrum at the electroweak 
scale. In particular, one of the Higgs-mass-squared parameters is driven 
to a negative value, ensuring that the electroweak symmetry is 
spontaneously broken.  The requirement that the electroweak symmetry be 
broken at precisely the right scale leads to a further constraint, which 
effectively fixes the magnitude of $\mu$ in terms of the other 
parameters, though its sign is still indeterminate. This version of the 
mSUGRA model, which depends on four parameters and a sign, viz. $\{m_0, 
m_{1/2}, A_0, \tan\beta, {\rm sgn}~\mu\}$, is called the cMSSM. Being 
more constrained, this model is also more predictive and hence is more 
readily testable. There exists, therefore, a vast amount of literature on this 
model, which has been studied with regard to ($a$)~collider signals 
\cite{Berger:2008cq, *Edsjo:2009rr, *Ozturk:2009fj, 
*Bhattacherjee:2009jh, *Baer:2009ff, *Genest:2009sv, *Feldman:2010uv, 
*Dreiner:2010gv}, ($b$)~low-energy processes, such as decays of $K, D$, 
and $B$ mesons \cite{Djouadi:2001yk, *Ellis:2002rp, *Gomez:2002tj, 
*Baer:2002gm, *Djouadi:2006be, *Heinemeyer:2008fb, Buchmueller:2009fn, 
*Buchmueller:2010ai}, and ($c$)dark-matter constraints arising from the 
fact that the relic density of the lightest SUSY particle (LSP) can be 
identified with the dark-matter content of the Universe 
\cite{Drees:1992am, *Jungman:1995df, *Nath:1997qm, *Barger:1997kb, 
*Bertone:2004pz, *Baer:2006dz, *Chattopadhyay:2009fr, 
*Bhattacharya:2009ij}. In this article, therefore, we shall focus on 
this model, though a simple extension will also figure into our analysis.

It is now common knowledge that null results from direct searches have 
pushed up the masses of sparticles into the regime of 100~GeV or above. 
However SUSY models can still make substantial contributions to 
low-energy processes, particularly those which are mediated by weak 
interactions. Among these, flavor-changing neutral current (FCNC) 
processes, with the famed Glashow-Iliopoulos-Maiani (GIM) cancellation, 
constitute a favored ground to look for SUSY effects (or any NP effects, 
for that matter). However, barring a few little hiccups, the SM rules 
supreme in the area of flavor physics, leaving very little room for NP 
theories, including SUSY and the cMSSM. Year by year, as the 
measurements of the FCNC processes grow better and better, the lower 
bounds on masses of new particles (including sparticles) have been 
creeping further and further up in order to squeeze the NP contributions 
into the ever-narrowing band of experimental errors in these 
measurements.

In this article, we consider one such recent low-energy experimental 
result, viz., the measurement of the branching ratio $B^+ \to \tau^+ 
\nu_\tau$. It directly constrains all models with minimal flavor 
violation (MFV), viz., models where all flavor-changing transitions are 
entirely governed by the Cabibbo-Kobayashi-Maskawa (CKM) matrix with no 
new phases beyond the CKM phase $\delta$. We find constraints on general 
NP with MFV that involves a charged Higgs boson. As the cMSSM (and 
almost any viable SUSY model) belongs to this category, we apply these 
constraints to the cMSSM and find a rather dramatic impact on the 
parameter space of the model. It turns out when we combine the results 
of the measurement in question with other low-energy measurements, such 
as the anomalous magnetic moment of the muon, and radiative and leptonic 
$B$ decays, most of the cMSSM parameter space is disfavored at the 
$95\%$ confidence level (C.L.). What survives all the constraints is a 
small patch in the four-dimensional parameter space $(m_0, m_{1/2}, A_0, 
\tan\beta)$ of the model, for a positive sign of $\mu$. This is very 
different from the kind of constraints derived from earlier, less 
restrictive measurements of $B^+ \to \tau^+ \nu_\tau$, where wide areas 
of the cMSSM parameter space were allowed.

As mentioned above, one of the attractive features of SUSY is that it 
provides a dark-matter candidate, viz., the lightest supersymmetric 
particle (LSP). This carries a conserved quantum number (R parity) which 
forbids its decay\footnote{Once again, this is not written in stone, for 
R-parity violation can happen and has been extensively studied 
\cite{Bhattacharyya:1997vv, *Dreiner:1997uz, *Chen:2010ss}.}. One can, 
therefore, study the evolution of the Universe in a SUSY model, and 
check whether the relic density of LSP's matches with the observed 
density of dark-matter as indicated by the cosmic microwave background 
radiation (CMBR) data \cite{Komatsu:2010fb}. Obviously, this matching 
will happen for only a small part of the parameter space of the model. 
It is encouraging that the dark-matter-allowed region in the cMSSM 
overlaps the small patch allowed by low-energy measurements quite 
substantially. We can say, therefore, that there exists a rather 
specific set of parameters which is simultaneously consistent with the 
low-energy data as well as with the hypothesis that LSP's form the dark-
matter content of the Universe. With this specific set of parameters, we 
generate the mass spectrum of sparticles, and find reasonably 
unequivocal indications as to the kind of signals expected at the LHC. 
No detailed analysis is necessary at this stage, for the relevant 
signals have already been considered in comprehensive studies by the 
ATLAS and CMS Collaborations \cite{ATLASnote,CMSnote}. Comparing their 
results with our parameter choice, we find that the 7~TeV run of the LHC 
may provide a weak indication of SUSY \cite{Baer:2010tk, 
*Bhattacharyya:2010gm, *Altunkaynak:2010we, *Chen:2010kq, 
*Bornhauser:2010mw}, which could be verified comprehensively even in the 
very early stages of the 14~TeV run. Going further, we may even say that 
if SUSY is indeed the correct NP option, then the LHC may eventually 
turn out to be the hoped-for SUSY factory, claimed in the literature 
\cite{Kane:1900zz}.
  
This article is organized as follows. In Sec.II we discuss the 
recent bounds on $B^+ \to \tau^+ \nu_\tau$ and how they affect MFV 
models. This is followed by Sec.III, where we discuss other low-energy measurements 
which constrain the cMSSM parameter space. The combined constraints are 
displayed and discussed in Sec.IV, where we also discuss the possible 
LHC signals which could arise therefrom. Sec.V discusses the 
so-called nonuniversal Higgs-mass (NUHM) model, a variant of the cMSSM, 
and some of its consequences. A critical summary of our results forms 
the substance of the concluding Sec.VI.


\section{The decay $B^+ \to \tau^+ \nu_\tau$}

On purely theoretical grounds, the leptonic decay $B^+ \to \tau^+ 
\nu_\tau$ is a clean decay mode, since the final state consists only of 
leptons and hence the usually troublesome strong rescattering phases are 
absent. Indeed, in the SM, the branching ratio of $B^+ \to \tau^+ 
\nu_\tau$ is given by the tree-level formula
\be
\mathrm{BR}(B^+ \to\tau^+\nu_\tau)_\mathrm{SM} = \frac{G_F^2 m_B 
m_\tau^2}{8\pi} \left( 1 - \frac{m_\tau^2}{m_B^2} \right)^2 \, f_B^2 \, 
|V_{ub}|^2 \, \tau_B \ ,
\label{BtaunuSM}
\ee
where $G_F$ is the Fermi constant, $\tau_{B}$ is the $B^+$ lifetime, 
$f_B$ = 192.8 $\pm$ 9.9 MeV \cite{Laiho:2009eu} is the $B^+$ decay 
constant, and $m_B, m_\tau$ are the masses of $B^+,\tau^+$, 
respectively. Here
\be
|V_{ub}| = (3.52 \pm 0.11) \times 10^{-3}
\label{eq:Vub}
\ee
is the relevant CKM matrix element, obtained through the combined fit 
\cite{Bona:2009cj, Charles:2004jd} to all the data excluding the $B^+ 
\to \tau^+ \nu_\tau$ measurements. The SM prediction, including 
higher-order corrections, is
\be
\mathrm{BR}(B^+ \to\tau^+\nu_\tau)_\mathrm{SM} 
= (0.81\pm 0.15)\times 10^{-4} \; .
\label{BtaunuSMno}
\ee
As recently as 2008, the experimental 
average value of this parameter \cite{arXiv:0808.1297(hep-ex)} was
\be
\mathrm{BR}(B^+ \to\tau^+\nu_\tau)_\mathrm{2008} 
= (1.41\pm 0.43)\times 10^{-4} \; ,
\label{BtaunuExpOld}
\ee
which was just about consistent with Eq.~(\ref{BtaunuSMno}) at 1 
standard deviation. At that time, it was shown 
\cite{Mahmoudi:2008ku} that corresponding constraints on the parameter 
space of the cMSSM (such as we discuss in this work) are rather minor.

Very recently (2010), however, new measurements of the branching ratio 
$\mathrm{BR}(B^+ \to\tau^+\nu_\tau)$ from B factories have changed the 
experimental value quite significantly. The most recent experimental 
measurements are \cite{BaBar:2008gx,BaBAr:2010rt,Hara:2010dk,Ikado:2006un}
\bea
\mathrm{Babar~ (semileptonic~ tag)}: &
\mathrm{BR}(B^+ \to \tau^+ \nu_\tau) = &
\left( 1.70 \pm 0.82 \right) \times 10^{-4} \; , \nonumber \\
\mathrm{(hadronic~ tag):} &  
\mathrm{BR}(B^+ \to \tau^+ \nu_\tau) = &
\left( 1.80 \pm 0.61 \right) \times 10^{-4} \; , \nonumber \\
\mathrm{Belle~ (semileptonic~ tag):} &  
\mathrm{BR}(B^+ \to \tau^+ \nu_\tau) = &
\left( 1.54 \pm 0.48 \right) \times 10^{-4} \; , \nonumber \\
\mathrm{(hadronic~ tag):} &  
\mathrm{BR}(B^+ \to \tau^+ \nu_\tau) = &
\left( 1.79 \pm 0.71 \right) \times 10^{-4} \; .
\eea
These results are quite consistent with each other. Combining these 
measurements, one gets the world average
\cite{ICHEP_T'JAMPENS}
\bea
\mathrm{BR}(B^+ \to \tau^+ \nu_\tau)_\mathrm{exp} 
= (1.68 \pm 0.31) \times 10^{-4} \; .
\label{Btaunuex}
\eea
Clearly, this measurement deviates significantly from the SM prediction 
given in Eq.~(\ref{BtaunuSMno}). Defining 
$R_{\tau\nu_\tau}^{\mathrm{exp}}$ to be \cite{Isidori:2006pk, 
Ellis:2007fu}
\begin{equation}
R_{\tau\nu_\tau}^{\mathrm{exp}}\equiv \frac{\mathrm{BR}(B^+\to\tau^+ 
\nu_\tau)_{\mathrm{exp}} }{\mathrm{BR}(B^+\to \tau^+ 
\nu_\tau)_{\mathrm{SM}}} \; ,
\end{equation}
and using Eqs.~(\ref{BtaunuSMno}) and (\ref{Btaunuex}),  we get  
\be
R_{\tau\nu_\tau}^{\mathrm{exp}} = 2.07 \pm 0.54 \; ,
\ee
which indicates a $\sim 2\sigma$ deviation. Deviations at this level 
frequently arise from statistical fluctuations in small data samples, or 
from the use of ill-determined theoretical quantities, and often 
disappear when more data are analyzed, or when more rigorous 
calculations are performed. In this case, however, the mismatch may not 
disappear so easily. For neither are the current measurements of 
$\mathrm{BR}(B^+ \to\tau^+\nu_\tau)$ based on a small statistics sample 
(see Refs. \cite{BaBar:2008gx,BaBAr:2010rt, Hara:2010dk, 
Ikado:2006un}), nor should one expect the SM prediction to change much, 
since the formula in Eq.~(\ref{BtaunuSM}) involves quantities that are 
already known pretty accurately. It appears, therefore, that it is 
sensible to at least explore the ability of NP beyond the SM to resolve 
the observed discrepancy.

Following Ref.~\cite{Isidori:2006pk}, we characterize the NP models that 
could potentially explain this anomaly by a quantity 
$R^{\mathrm{NP}}_{\tau\nu_\tau}$, defined as
\begin{equation}
R^{\mathrm{NP}}_{\tau\nu_\tau}\equiv\frac{\mathrm{BR}(B^+\to\tau^+ 
\nu_\tau) _{\mathrm{\,SM+NP}} } 
{\mathrm{BR}(B^+\to\tau^+\nu_\tau)_{\mathrm{SM}}} \; ,
\end{equation}
where the subscript SM+NP represents the net branching ratio in the NP 
scenario, including the SM contribution. The 95$\%$ C.L. allowed range 
for $R^{\mathrm{NP}}_{\tau\nu_\tau}$ then works out to
\begin{equation}
0.99 < R^{\rm{NP}}_{\tau\nu_\tau} < 3.14 \, ,
\label{R-bounds}
\end{equation}
which essentially means that NP models with positive contributions are 
favored by the data and those with negative contributions are quite 
strongly disfavored.

There exist, of course, a wide variety of models of NP which could 
provide extra contributions to the branching ratio of $B^+ \to \tau^+ 
\nu_\tau$. However, we focus only on the MFV models. For a large class 
of MFV models that involve a charged Higgs boson $H^+$ -- such as 
two-Higgs doublet models, of which the cMSSM is an example -- the 
branching ratio of $B^+ \to \tau^+ \nu_\tau$ is given by 
\cite{Hou:1992sy}
\be
\mathrm{BR}(B^+ \to\tau^+ \nu_\tau)_\mathrm{NP}=
\frac{G_F^2 m_B m_\tau^2}{8\pi}
\left( 1 - \frac{m_\tau^2}{m_B^2} \right)^2
f_B^2 \, |\widetilde{V}_{ub}|^2 \, \tau_B
\left(1-\tan^2\beta \frac{m^2_B}{M^2_{+}} \right)^2 \; 
\label{BtaunuNP}
\ee
at the tree level, where $M_+$ is the mass of the charged Higgs boson. 
Here NP stands specifically for the MFV model, but we retain the 
notation ``NP`` in the interests of simplicity\footnote{
Note that our analysis for the MFV models in this section
closely follows that of \cite{Bona:2009cj}, with minor differences. 
Our constraints in the $M_+$--$\tan\beta$ parameter space naturally 
are almost identical. However, we present the detailed analysis here 
for the sake of completeness and clarification of our procedure.
}.

In the above formula $|\widetilde{V}_{ub}|$ is the value of $|V_{ub}|$ 
obtained in the context of the NP model, which in general will be 
different from $|V_{ub}|$ obtained from the data in the context of the 
SM. In order to get rid of this uncertainty in the CKM parameter, we 
restrict ourselves to the determination of $|V_{ub}|$ through only those 
measurements that are independent of NP. Such a fit is called the fit to 
the universal unitarity triangle (UUTfit) \cite{Buras:2000dm}, and it 
uses only
\begin{itemize}
\item the measurements of $|V_{ub}/V_{cb}|$ from semileptonic $B$ 
decays,
\item the ratio of mass differences in the $B_s$ and $B_d$ systems: 
$|\Delta M_s/\Delta M_d|$, and
\item the measurement of $\sin 2\beta$ from the time-dependent $CP$ 
asymmetry in $B_d \to J/\psi K^{(*)}$.
\end{itemize}
The UUTfit value of $|V_{ub}|$ comes out as \cite{Bona:2009cj}
\begin{equation}
|V_{ub}|_{\mathrm{UUTfit}} = (3.50 \pm 0.12) \times 10^{-3} \ ,
\label{Vub}
\end{equation}
which is actually very close to the global fit in Eq.~(\ref{eq:Vub}). 
Using this value, the SM prediction for the branching ratio of $B^+ \to 
\tau^+ \nu_\tau$ changes slightly from Eq.~(\ref{BtaunuSMno}) and 
becomes
\be
\mathrm{BR}(B^+ \to\tau^+\nu_\tau)_\mathrm{SM} 
= (0.80\pm 0.15)\times 10^{-4} \; .
\label{BtaunuSM_UTfit}
\ee
Note that while the UUTfit \cite{Bona:2009cj} is obtained using 
the lattice prediction $f_B = 200 \pm 20$ MeV \cite{Lubicz:2008am}
of the LQCD Collaboration,
we use the more recent, averaged value from lattice simulations,
$f_B$ = 192.8 $\pm$ 9.9 MeV \cite{Laiho:2009eu},
which has a much smaller error\footnote{
Ideally, of course, the UUTfit needs to be performed again with the updated
$f_B$ value. We have assumed that the updated fit will not 
significantly affect the $|V_{ub}|$ value.
}, for the calculation of
$\mathrm{BR}(B^+ \to\tau^+\nu_\tau)_\mathrm{SM}$. 
The 95\% C.L. allowed range for $R^{\mathrm{NP}}_{\tau\nu_\tau}$ assumes 
the value
\begin{equation}
0.99 < R^{\rm{NP}}_{\tau\nu_\tau} < 3.19 \, ,
\label{R-bounds_UUTfit}
\end{equation} 
which forms the basis of all subsequent analyses in this article. Once 
$|V_{ub}|$ is chosen in this ''model-independent'' way, we can take 
$|\widetilde{V}_{ub}| = |V_{ub}|_{\mathrm{UUTfit}}$, and hence the 
theoretical MFV prediction for $R^{\mathrm{NP}}_{\tau\nu}$ at the tree 
level becomes
\be
R^{\mathrm{NP}}_{\tau\nu_\tau}|_{\rm tree} = 
\left(1-\tan^2\beta \frac{m^2_B}{M^2_{+}} \right)^2 \; . 
\ee 
If higher-order corrections are included then this ratio gets modified 
\cite{Akeroyd:2003zr} to a form
\be
R^{\mathrm{NP}}_{\tau\nu_\tau} = 
\left(1-\frac{\tan^2\beta}{1+\tilde\epsilon_0\,\tan\beta}\,\frac{m^2_B} 
{M^2_{+}}\right)^2 \; ,
\label{R-NP-HO}
\ee
where $\tilde\epsilon_0$ encodes all the higher-order corrections, 
which, of course, will have some dependence on the free parameters of 
the MFV model. We take the range of $\tilde\epsilon_0$ to be $-0.01 \leq 
\tilde\epsilon_0 \leq 0.01$, as obtained in \cite{Buras:2002vd} by a 
scan over reasonable values of the MFV model parameters. When a 
specific model, such as the cMSSM, is considered, $\tilde\epsilon_0$ can 
be calculated explicitly.

\begin{figure*}[ht]
\centerline{\epsfig{file=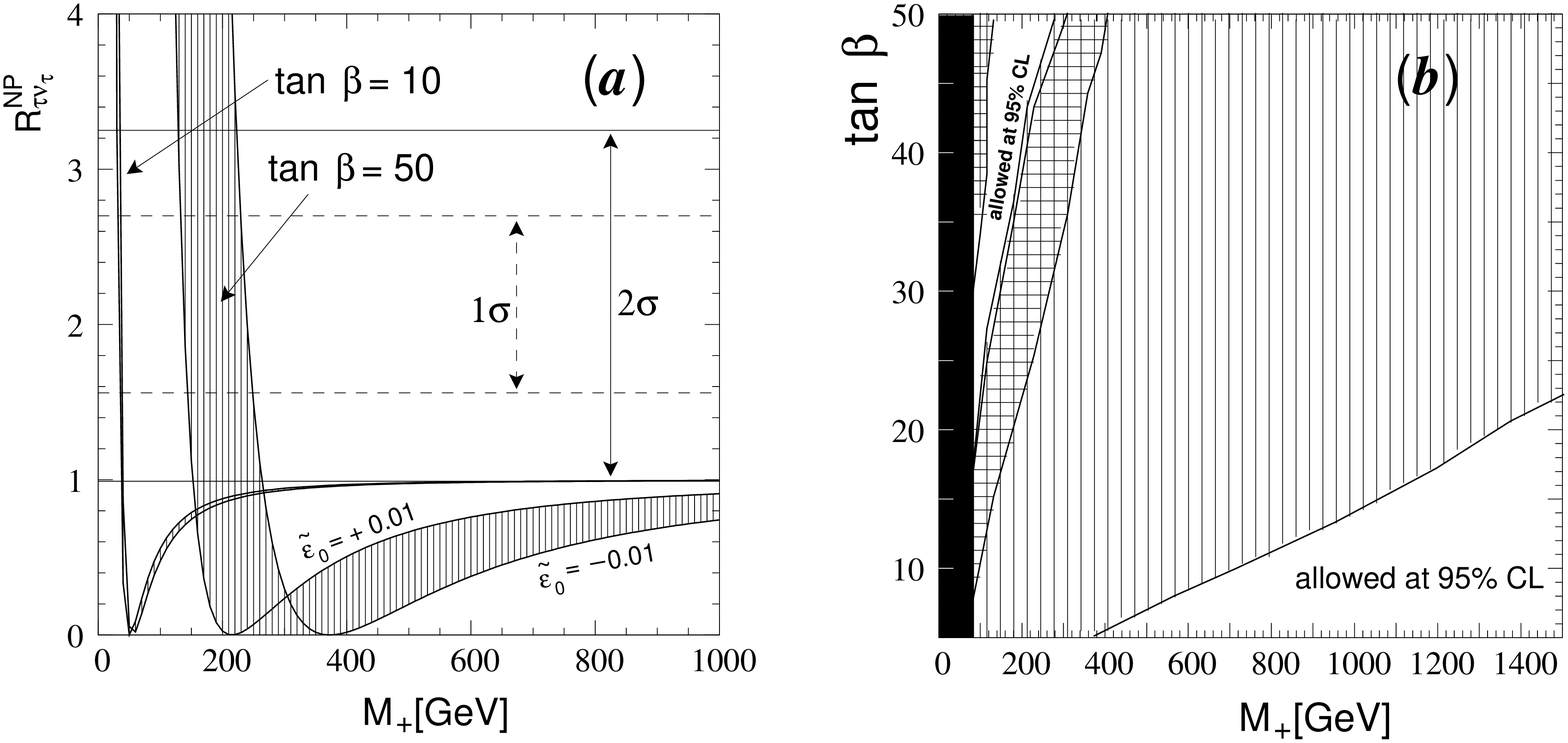,height=6.5cm,width=13.5cm}}
\caption{\small ($a$) The dependence of $R^{NP}_{\tau\nu_\tau}$ 
on the mass $M_+$ of the charged Higgs boson in MFV models for two 
values $\tan\beta = 10$ and 50, and ($b$) the 95\% C.L. constraints on 
the $M_+$--$\tan\beta$ plane. The vertically hatched regions in ($a$) 
correspond to higher order corrections varying between $\tilde\epsilon_0 
= -0.01$ and $+0.01$, while the $1\sigma$ ($2\sigma$) experimental 
measurements of $R^{NP}_{\tau\nu_\tau}$ are shown by horizontal broken 
(solid) lines. The dark band in ($b$) corresponds to the LEP bound. The 
large, vertically hatched region in ($b$) is disallowed by the recent 
(2010) $R^{NP}_{\tau\nu_\tau}$ constraint, while the 
horizontally hatched region is disallowed by the 2008 data.}
\label{B_tau_nu}
\end{figure*}

The impact of the experimental data on MFV models with a charged Higgs 
boson, as discussed above, can be clearly discerned from 
Fig.~\ref{B_tau_nu}($a$), where we plot the value of 
$R^{NP}_{\tau\nu_\tau}$ as a function of the charged Higgs boson mass 
$M_+$. As Eq.~(\ref{R-bounds_UUTfit}) indicates, such a model should 
tend to make $R^{\mathrm NP}_{\tau \nu_\tau}$ greater than unity, and 
there is very little room for $R^{\mathrm NP}_{\tau \nu_\tau} < 1$. 
However, the negative sign on the right side of Eq.~(\ref{R-NP-HO}) 
indicates that unless the NP contribution is very large, the models in 
question have a tendency to diminish $R^{\mathrm NP}_{\tau \nu_\tau}$ 
rather than augment its value.  As a result, a model with a heavy 
charged Higgs boson {\it cannot} be considered as an explanation for the 
deviation of ${\rm BR}(B^+ \to \tau^+ \nu_\tau)$ from its SM value. 
Instead, if we do have such a model, we would expect rather strong 
constraints on its parameters, since the NP contribution must be 
squeezed into the small tolerance below unity, as given in 
Eq.~(\ref{R-bounds_UUTfit}). Such a situation would naturally arise for 
large $M_+$, when the NP contribution becomes negligible, and 
$R^{NP}_{\tau\nu_\tau} \to 1$ for all $\tan\beta$ -- though it always 
stays less than unity. This corresponds to the rising part of the 
curves, towards the right end of Fig.~\ref{B_tau_nu}($a$). A glance at 
the figure will, however, leave no doubt that this limiting case is 
barely allowed at $2\sigma$ for low $\tan\beta$ ( = 10), but disallowed 
for high $\tan\beta$ (= 50). We surmise, therefore, that for high values 
of $M_+$, the $B^+ \to \tau^+ \nu_\tau$ measurement favors low values of 
$\tan\beta$.

For low values of $M_+$, on the other hand, Eq.~(\ref{R-NP-HO}) tells us 
that it is possible for the NP contribution to be so large that it 
completely dominates the SM contribution, and in this limit, it is 
possible to have $R^{NP}_{\tau\nu_\tau} \gtrsim 1$. The explicit 
condition is
\begin{equation}
\frac{\tan^2\beta}{1 + \tilde\epsilon_0 \tan\beta} \gtrsim \frac{2 
M_+^2}{m_B^2} \ .
\end{equation}
However this can happen only for a very restricted set of 
$(M_+,\tan\beta)$ values, some of which are already constrained 
experimentally. For example, for $\tan\beta=10$, one can have 
$R^{NP}_{\tau\nu_\tau}$ well inside the $2\sigma$ range only if $M_+$ is 
$\lesssim 50$ GeV, but such low $M_+$ values are ruled out by the LEP 
direct searches, which give $M_+ > 79.3$ GeV \cite{Amsler:2008zzb}. On 
the other hand, if $\tan\beta=50$, the same LEP data permit 
140~GeV$<M_+<$~220~GeV which can render $R^{NP}_{\tau\nu}$ well inside 
the $2\sigma$ range, as shown in Fig.~\ref{B_tau_nu}($a$). Thus, one may 
complement our earlier assertion by the statement that in the opposite 
limit, i.e. for low values of $M_+$, the $B^+ \to \tau^+ \nu_\tau$ 
measurement favors high values of $\tan\beta$. This is also the limit in 
which the NP models contribute positively in accounting for the 
deviation of the experimental data from the SM predictions.

The two limits are made explicit in Fig.~\ref{B_tau_nu}($b$), which 
shows the 95$\%$ C.L. constraints on the $M_+$--$\tan\beta$ plane. The 
dark band represents the LEP constraint $M_+ > 79.3$~GeV and the 
vertically-hatched region is disallowed by the $B^+ \to \tau^+ \nu_\tau$ 
measurement. This leaves only two small unshaded regions for low $M_+$ 
and high $M_+$, in accordance with the above discussion. Our result may 
be contrasted with the constraints obtained from the 2008 data, which 
are shown by horizontal hatching, and constrain only a small region with 
$M_+ < 400$~GeV and somewhat high $\tan\beta$. One may say, therefore, 
that the recent measurement of the branching ratio for $B^+ \to \tau^+ 
\nu_\tau$ has considerably improved the constraints on MFV models with 
charged Higgs bosons. As the cMSSM belongs to this category, we should 
expect correspondingly severe constraints on the corresponding parameter 
space when we compare its predictions with this new experimental result.


\section{Other constraints}

When we consider an all-encompassing model like the cMSSM, with 
far-flung implications in almost all areas of electroweak physics, the 
constraints arising from $B^+ \to \tau^+ \nu_\tau$ cannot be considered 
in isolation, but must be combined with other bounds -- some of which 
are equally restrictive, at least at the $2\sigma$ level. These 
constraints can be classified as follows.
\begin{enumerate}
\item {\it Theoretical constraints}.-- arising from requirements of 
internal consistency of the model. In particular, if the Higgs-mass 
parameter which should be driven negative by RG running remains 
positive, we cannot explain electroweak symmetry breaking (EWSB) in this 
model. There is also a substantial region where the model predicts that 
the charged stau $\widetilde{\tau}_1$ is the LSP, and is therefore 
precluded by the absence of a large relic density of charged particles.
\item {\it Collider bounds}.-- arising from the nondiscovery in direct 
searches \cite{Amsler:2008zzb} at the CERN LEP and Fermilab Tevatron of 
predicted particles, most notably the light Higgs boson $h^0$ and the 
lighter chargino $\widetilde{\chi}^+_1$. While the chargino couplings 
are large enough for the experimental bounds to practically saturate the 
kinematic reach of these machines, the light Higgs boson mass in SUSY 
models is generally sensitive to higher-order corrections, where there 
is a theoretical uncertainty of around 3--4~GeV at the 
next-to-next-to-leading order (NNLO) and higher \cite{Heinemeyer}. To 
take care of this, we consider a softer lower bound of 111~GeV, rather 
than the kinematic bound of 114.4~GeV usually applied to the SM Higgs 
boson.
\item {\it Indirect bounds}.--arising from measurements of low-energy 
processes where new particles and interactions in NP models can also 
contribute. In the context of the cMSSM, the most important of these are 
the measurements of ($a$)~the anomalous magnetic moment of the muon, 
($b$)~the rate of the radiative decay $B_d \to X_s \gamma$, and ($c$)~the 
BR for the leptonic decay $B_s \to \mu^+ \mu^-$. 
Here we have assumed that the NP is of the MFV kind, and that it
survives the measurements other than those explicitly mentioned above.
In particular, the large $B_s$--$\overline{B}_s$ mixing phase, 
or the $A_{CP}(B \to K \pi)$ measurements, cannot be explained by
any MFV models, and we assume that these anomalies will disappear
with more data or with better theoretical calculations. 
\end{enumerate} 
Of the above, the theoretical and direct search constraints may be 
considered firm constraints, as they are unlikely to be changed by 
inclusion of further types of NP along with the cMSSM or whatever model 
is being studied. On the other hand, constraints from indirect 
measurements are not so robust, as they can easily change if some new 
effect is postulated.  Before we proceed to apply these constraints to 
the cMSSM parameter space, therefore, a brief discussion of the actual 
measurements used in our analysis is called for. This forms the 
remaining part of this section.

\begin{itemize}
\item {\it The anomalous magnetic moment of the muon, $a_{\mu}=(g-2)/2$:} 
This is one of the most compelling indicators of NP and it is well known 
as a major constraint for NP theories such as supersymmetry or extra 
dimensions. The latest measured value \cite{Bennett:2006fi} for 
$a_{\mu}$ is
\bea
a_{\mu}^{\textnormal{exp}}= (11659208.0 \pm 6.3) 
\times 10^{-10} \; .
\eea
In the recent past, the SM prediction \cite{Buras:2000dm} for $a_\mu$ 
has undergone numerous vicissitudes with respect to the experimental 
data, occasionally being consistent with it and occasionally deviating 
at the level of 2$\sigma$-3$\sigma$. Much of the difficulty in making this 
prediction accurate lies in the fact that the experimental measurement 
is sensitive to two-loop corrections where some nonperturbative QCD 
effects due to the low mass scale are involved. The latter have to be 
obtained by fitting experimental data, whose errors then feed into the 
theoretical uncertainty. The most recent SM prediction is 
\cite{Miller:2007kk}
\bea
a_{\mu}^{\textnormal{SM}}= (11659178.5 \pm 6.1) 
\times 10^{-10} .
\eea
The discrepancy between the SM and experiment is, therefore,
\bea
{\Delta a}_{\mu}^{\textnormal{exp}} = 
a_{\mu}^{\textnormal{exp}} -a_{\mu}^{\textnormal{SM}}= 
(29.5 \pm 8.8) \times 10^{-10} \; .
\eea
This is at the somewhat high level of $\sim 3.4 \sigma$, but is not 
normally considered a ``smoking gun`` signal for NP for reasons stated 
above. Nevertheless, in order to check if this discrepancy {\it can} be 
explained with the cMSSM, we use a procedure \cite{Eriksson:2008cx} that 
does not calculate the two-loop SUSY corrections, but includes them in 
the theoretical errors, to obtain a 95\% C.L. range
\bea
11.5 \times 10^{-10} < \Delta a_{\mu}^{\rm NP} < 47.5 \times 10^{-10} \; ,
\label{Damu}
\eea
where $\Delta a_{\mu}^{\rm NP}$ is the extra contribution due to NP. In 
our analysis, the NP in question will be the cMSSM, or a variant, but we 
choose, as in the previous section, to retain the label ''NP``.

Obviously, in the cMSSM, the value of $\Delta a_{\mu}^{\rm NP}$ will 
depend on all the free parameters of the model. However, it is known 
that the sign of the cMSSM contribution is directly sensitive to the 
sign of the $\mu$ parameter \cite{Chattopadhyay:1995ae, 
*Chattopadhyay:2001wt}: for $\mu < 0$, the cMSSM contribution is 
negative, while for $\mu > 0$, a positive contribution is predicted by 
some regions of the cMSSM parameter space. Since the 95\% C.L. range of 
$\Delta a_{\mu}^{\rm NP}$ indicated in Eq.~(\ref{Damu}) is entirely 
positive, it indicates that the sign $\mu < 0$ is disallowed by the 
measurement of the muon anomalous magnetic moment, and even with $\mu > 
0$, there are strong constraints on the remaining parameters of the 
cMSSM.

\item {\it The radiative decay $B_d \to X_s \gamma$:} In the SM, the BR 
of the radiative decay $B_d \to \textnormal{X}_s \gamma$ has been 
calculated \cite{Lunghi:2006hc, *Misiak:2006ab, *Misiak:2006zs, *Freitas:2008vh}
to NNLO in QCD to be
\bea
\textnormal{BR}(\textnormal{B}_d \to \textnormal{X}_s 
\gamma)_{\textnormal{SM}}
= (3.15 \pm 0.23) \times 10^{-4} \; .
\eea
The current experimental average for the BR by the Heavy Flavor 
Averaging Group (HFAG)~\cite{HFAG_btosg} is 
\bea
\textnormal{BR}(\textnormal{B}_d \to \textnormal{X}_s 
\gamma)_{\textnormal{exp}} 
= \left( 3.55 \pm 0.26 \right) \times 10^{-4} \; ,
\eea
which is consistent with the SM prediction within 1 standard 
deviation, leaving very little room for NP contributions. As a result, 
this measurement has a tremendous impact on MFV models involving a 
charged Higgs boson $H^+$, essentially pushing up the mass $M_+$ to very 
large values. Of all such models, the constraints on SUSY models can be 
more relaxed because of large cancellations between the charged Higgs 
boson contributions and the chargino contributions which are the 
hallmark of SUSY models. Nevertheless, there do exist bounds on the 
cMSSM arising from the residual contribution, especially for light 
$M_+$, and these have to be taken into consideration.

In our analysis, after including the theoretical uncertainties in the 
cMSSM following the method outlined in \cite{Ellis:2007fu}, we set the 
95\% C.L. range for the branching ratio to be
\bea
2.05 \times 10^{-4} \leq \textnormal{BR}(\textnormal{B}_d \to 
\textnormal{X}_s \gamma) \leq 5.05 \times 10^{-4} \; .
\eea
It turns out that the $\textnormal{B}_d \to \textnormal{X}_s \gamma$ 
constraint is also extremely sensitive to the sign of $\mu$. For 
$\mu<0$, it eliminates a large part of the parameter space 
\cite{Nath:1994tn}, while for $\mu>0$ the constraint is comparatively 
weaker. Neither of these constraints is as strong as those arising from 
the muon $(g-2)/2$; however, they are complementary to it.
 
\item {\it The leptonic decay $B_s \to \mu^+ \mu^-$:} Within the SM, the 
fully leptonic decay $B_s \to \mu^+ \mu^-$ is chirally suppressed; the 
SM prediction is
\bea
\textnormal{BR}(\textnormal{B}_s \to \mu^+ \mu^-)_{\textnormal{SM}} 
= (3.19 \pm 0.35) \times 10^{-9} \; .
\eea
The uncertainty in the BR comes principally from the decay constant 
$f_{B_s}= 238.8 \pm 9.5$ MeV \cite{Laiho:2009eu} and from the CKM 
element $|V_{ts}| = 0.041 \pm 0.001$ \cite{Amsler:2008zzb}. The current 
experimental upper bound by the CDF Collaboration is \cite{CDF9892}
\bea
\textnormal{BR}(\textnormal{B}_s \to \mu^+ \mu^-)_{\textnormal{CDF}} 
< 4.3 \times 10^{-8}~ (95\% ~ \textnormal{C.L.}) \; .
\eea
After including the theoretical uncertainties, we get the 95$\%$ C.L. 
upper limit 
\bea
\textnormal{BR}(\textnormal{B}_s \to \mu^+ \mu^-) < 4.8 \times 10^{-8}.
\eea
Inclusion of charged Higgs bosons, whose left- and right-chiral 
couplings depend on $\cot\beta$ and $\tan\beta$, respectively, has a 
direct impact on the BR for $B_s \to \mu^+ \mu^-$, which gets enhanced 
considerably above the SM prediction, and easily saturates the upper 
bound, especially for low values of $M_+$ and large $\tan\beta$. Indeed, 
for large $\tan\beta$, the cMSSM contribution is known 
\cite{Lunghi:2006uf} to scale as $\tan^6 \beta/(M_+^2 - M_W^2)^2$. Thus, 
this process also constrains MFV models with a charged Higgs boson. It 
turns out that for the cMSSM, these constraints are not more severe than 
the combination of all other constraints; however, we shall demonstrate 
later that they do have an impact if the assumptions of the cMSSM are 
relaxed.

\end{itemize}

In this section, we have listed the major constraints, apart from the 
new data on $B^+ \to \tau^+ \nu_\tau$, on the parameter space of SUSY 
models, of which the cMSSM will be showcased in the following section. 
It may be noted in passing that this list is not fixed for all time, as 
there are several other low-energy processes and direct search bounds 
which also constrain the SUSY parameter space. Current data on these 
rule out patches of the parameter space which are subsumed in the 
disallowed regions arising from the constraints which we have listed 
above. However, it is entirely possible that a future measurement -- 
including some LHC searches -- could rule out wider patches of the SUSY 
parameter space, and then the relevant processes would have to be taken 
into consideration. With this caveat, we now turn to the explicit 
constraints on the cMSSM parameter space, and the impact of the $B^+ \to 
\tau^+ \nu_\tau$ measurement on this analysis.


\section{Constraining the cMSSM} 

As mentioned in the Introduction, the parameter space of the cMSSM has 
four unknowns, viz. $m_0$, $m_{1/2}$, $A_0$, $\tan\beta$, that can take 
real values. The sign of the $\mu$ parameter is also undetermined, but 
as indicated in the previous section, the muon $(g - 2)/2$ constraint 
disfavors $\mu < 0$. We therefore restrict ourselves to $\mu > 0$ and, 
hence, consider a simply connected parameter space of four dimensions.

The theoretical ranges of the parameters $m_0$, $m_{1/2}$, and $A_0$ are, 
in principle, completely undetermined, but the region of interest is 
clearly that which would lead to sparticle masses kinematically 
accessible to current accelerators such as the LHC. Keeping this in 
mind, we scan the ranges
\begin{equation}
0 \leq m_0 \leq 2~\mathrm{TeV} \; ,
\qquad\qquad 0 \leq m_{1/2} \leq 1~\mathrm{TeV}\; ,
\qquad\qquad -2~\mathrm{TeV} \leq A_0 \leq 2~\mathrm{TeV} \; .
\end{equation}
The range of the remaining parameter $\tan\beta$ is determined mainly by 
its impact on the scalar Higgs sector of the cMSSM, where, indeed, it 
arises. For very low values of $\tan\beta$ ($\sim 1$), one tends to 
predict the lightest Higgs boson $h^0$ to have a small mass, which is 
already ruled out by the LEP constraints. On the other hand, if 
$\tan\beta > m_t/m_b$, the couplings of the charged Higgs boson to a 
$t\bar{b}$ pair begin to enter the nonperturbative regime. We have 
chosen, therefore, the reasonable range
\begin{equation}
4 \leq \tan\beta \leq 50 \; . 
\end{equation} 

Using these parameters, we perform a numerical scan over the cMSSM 
parameter space, using ($a$)~{\sc SuSpect} \cite{Djouadi:2002ze} to 
generate the mass spectrum (this also takes care of the theoretical and 
direct search constraints), ($b$)~{\sc SuperIso} \cite{Mahmoudi:2007vz, 
*Mahmoudi:2008tp} to calculate the variables listed as indirect 
constraints in the previous section, and, finally, ($c$)~{\sc 
micrOMEGAs} \cite{Belanger:2001fz, *Belanger:2006is} to calculate the 
dark-matter relic density. All of these are state-of-the-art software 
in the public domain, guaranteed to include higher-order corrections as 
available at the moment, which have been tested in multifarious 
applications, as the literature testifies.

\begin{figure*}[ht]
\centerline{ \epsfxsize= 14.0 cm \epsfysize= 7.5 cm \epsfbox{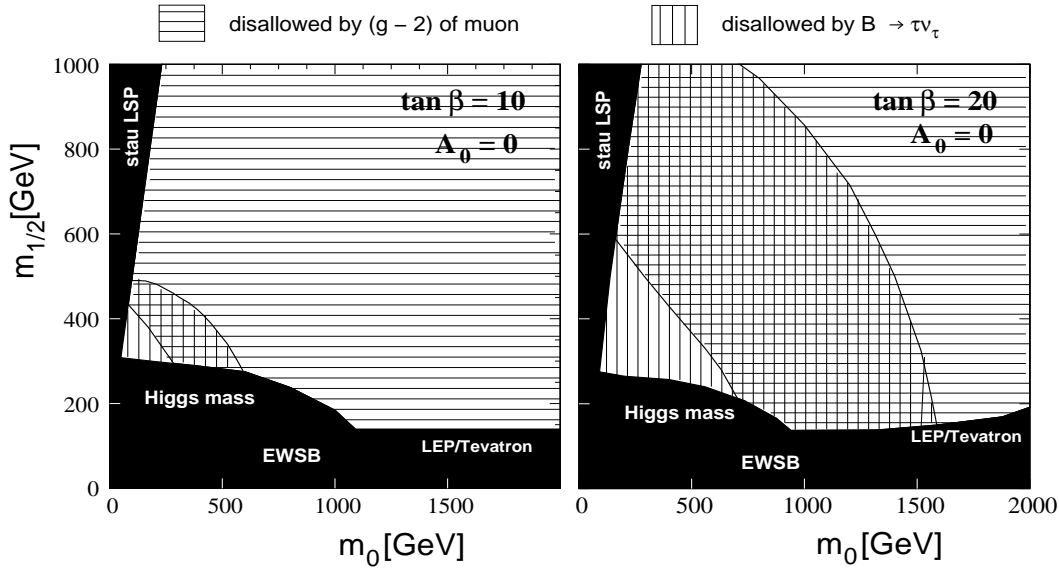} }
\caption{\small The 95\% C.L. constraints on the $m_0$--$m_{1/2}$ plane, for 
$A_0 = 0$ and $\tan\beta = 10$ (left panel) and $\tan\beta = 20$ (right 
panel). Horizontal (vertical) hatching indicates regions ruled out by 
the measurement of the muon $(g-2)/2$ [$BR(B^+ \to \tau^+ \nu_\tau)$] and 
the cross-hatched region represents their overlap. The constraint from 
$B \to X_s \gamma$ is subsumed in that from the lower bound on the Higgs 
mass from direct searches, and hence is invisible in these plots. We 
take $\mu > 0$ in this and all subsequent plots.}
\label{mSUGRA-1}
\end{figure*}

In Fig.~\ref{mSUGRA-1}, we present the constraints on the cMSSM 
parameter space in the $m_0$--$m_{1/2}$ plane, where the tension between 
various measurements appears quite clearly. For this figure, we have set 
$A_0 = 0$, which is an assumption commonly made for simplicity. We take 
$\mu > 0$ as required by the muon $(g-2)/2$ constraint. The panel on the 
left (right) corresponds to $\tan\beta = 10~(20)$. The dark areas are 
ruled out by the theoretical and direct search constraints explained in 
the previous section, with the approximate areas highlighted in white 
lettering\footnote{Following common practice, we do not delineate 
separate patches in the firmly disallowed (dark) region in detail, as 
that would not be germane to the present discussion.}. Focusing on the 
left panel, it is clear that the muon $(g-2)/2$ constraint (indicated by 
horizontal hatching) is very stringent, ruling out almost all the region 
considered, and allowing only a small patch with low values of $m_0$ and 
$m_{1/2}$. This patch, however, is disallowed by the new measurement of 
$B^+ \to \tau^+ \nu_\tau$, as can be seen from the vertical hatching. 
The overlap between the disallowed regions is indicated by 
cross-hatching.  It is quite clear, therefore, that for $\tan\beta = 
10$, there is {\it no} region in the parameter space shown that is at 
once consistent (to 95\% C.L.) with both the anomalous muon magnetic 
moment and the $B^+ \to \tau^+ \nu_\tau$ branching ratio. We have 
checked that even if we take $\tan\beta$ down to values as small as 
$\tan\beta = 4$, the two measurements, taken together with the firm 
constraints, do not allow for a simultaneously allowed parameter space. 
At higher value of $\tan\beta$, the situation is even worse. This is 
apparent from the right panel, where our results are plotted for 
$\tan\beta = 20$. Here it is true that a larger region is permitted by 
the muon $(g-2)/2$ measurement, but the region disallowed by the $B^+ 
\to \tau^+ \nu_\tau$ branching ratio is also much larger and covers the 
entire region allowed by $(g-2)/2$. The region disallowed by $B^+ \to 
\tau^+ \nu_\tau$ grows for larger value of $\tan\beta$, as may be 
guessed from Fig.~\ref{B_tau_nu}($b$), and for values of $\tan\beta \sim 
50$, it would cover the whole of the parameter space shown in the panels 
of Fig.~\ref{mSUGRA-1}. Thus, for $A_0 =0$, one may say that these two 
measurements alone are enough to ensure that the full mSUGRA parameter 
space is strongly disfavored.

It may be noted in passing that among the other constraints from the 
low-energy data, the patch disallowed by $B_d \to X_s \gamma$ is 
subsumed in that from the Higgs-mass bound, and, likewise, the patch 
inconsistent with $B_s \to \mu^+\mu^-$ is overlaid by the dark region 
corresponding to the firm constraints. We have not, therefore, shown 
these disallowed regions in Fig.~\ref{mSUGRA-1}.

\begin{figure*}[ht]
\centerline{ \epsfxsize= 14.0 cm \epsfysize= 8.0 cm \epsfbox{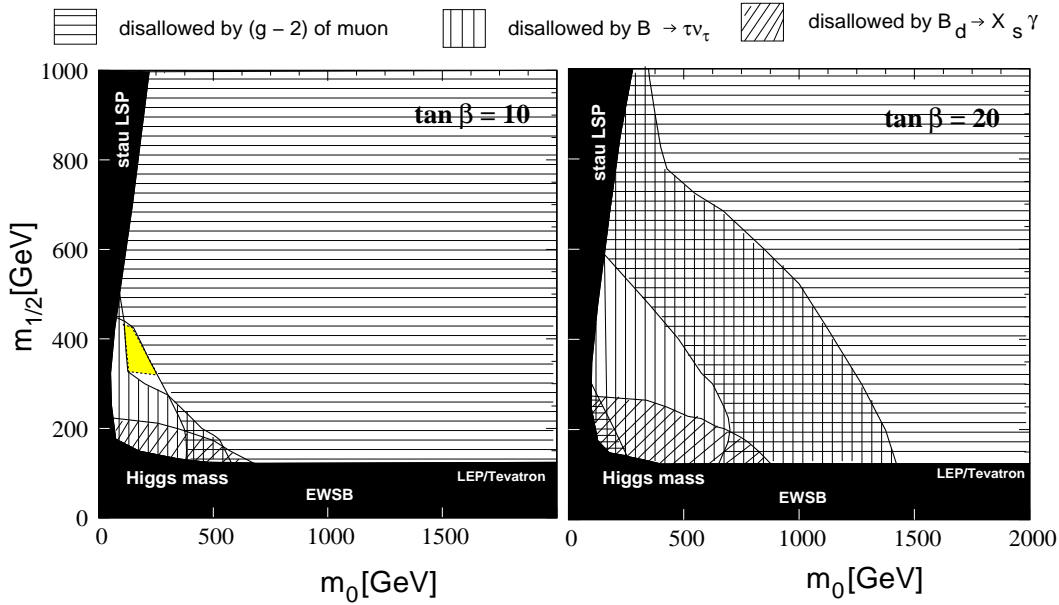} }
\caption{{\small The same as in Fig.~\ref{mSUGRA-1}, except that the 
trilinear coupling $A_0$ is kept floating over the range $(-2,+2)$~TeV, 
and regions are considered disallowed only if they remain disallowed for 
all values of $A_0$ in the given range. This weakens the constraints 
enough for a small allowed region to appear in the left panel 
($\tan\beta = 10$). The yellow/light gray region is simultaneously 
consistent with {\em all} constraints at 95\% C.L.}}
\label{mSUGRA-2}
\end{figure*}

The above result, disappointing as it may appear, is by no means the end 
of the road for the cMSSM, for it has been obtained only on the slice of 
parameter space for which $A_0 = 0$. The situation changes when we 
permit $A_0$ to vary. This affects the running of the charged Higgs 
boson mass, and we find that for large negative values of $A_0$, for a 
given $\tan\beta$, the Higgs-mass $M_+$ is driven to larger values than 
what one would obtain by setting $A_0 = 0$. In the context of 
Fig.~\ref{B_tau_nu}($b$), this effect then pushes the model horizontally 
in the $M_+$--$\tan\beta$ plane, eventually penetrating into the 
''allowed'' region. The $B^+ \to \tau^+ \nu_\tau$ constraint, therefore, 
can be quite considerably weakened by choosing large negative values of 
$A_0$. We do not expect such a significant change in the muon $(g-2)/2$ 
constraint, but some relaxation is not unreasonable to expect when $A_0$ 
is varied over a wide range. Large negative values of $A_0$ also tend to 
increase the mass of the lightest Higgs boson $h^0$, thereby relaxing 
somewhat the LEP bounds arising from Higgs boson mass considerations 
\cite{Carena:2002es}. Accordingly, we repeat our analysis of the 
constraints on the $m_0$--$m_{1/2}$ plane, keeping $A_0$ floating 
between $-2$~TeV and $+2$~TeV. In our analysis, a point in the 
$m_0$--$m_{1/2}$ plane, for a given $\tan\beta$, is taken to be allowed 
at 95\% C.L. by a given constraint if we can find any value of $A_0$, 
lying in the range $-2~\mathrm{TeV} \leq A_0 \leq 2~\mathrm{TeV}$, for 
which the given constraint is satisfied. Our results are exhibited in 
Fig.~\ref{mSUGRA-2}, which follows the notations and conventions of 
Fig.~\ref{mSUGRA-1} closely. In addition to the constraints shown 
therein, the one from $B_d \to X_s\gamma$ now makes its appearance as a 
small region hatched with slanting lines, indicating that this 
constraint is now stronger than the weakened Higgs-mass bound from 
collider machines. However, it is not strong enough to rule out any 
portion which is not already disallowed by the other constraints.

Even a cursory examination of the dark and hatched regions in 
Fig.~\ref{mSUGRA-2} will indicate that, while the qualitative features of 
the regions disfavored by the muon $(g-2)/2$ and the $B^+ \to \tau^+ 
\nu_\tau$ measurements stay the same, somewhat larger areas in the plane 
are ``allowed`` by each constraint individually. This, by itself, is not 
surprising, but it has the exciting consequence that now the left panel 
($\tan\beta = 10$) exhibits a small patch, roughly triangular in shape, 
which satisfies each of the constraints individually for {\it some} 
value of $A_0$, and moreover, there is a subset of this region where all 
the constraints are satisfied simultaneously for the {\it same} value of 
$A_0$. This subregion, which represents the actual parameter space 
consistent with all the measurements individually at 95\% C.L., is 
denoted by yellow/light gray shading. It is on this ''allowed'' region 
that we focus our interest in the subsequent discussion.

\begin{figure*}[ht]
\centerline{ \epsfxsize= 14.0 cm \epsfysize= 8.0 cm \epsfbox{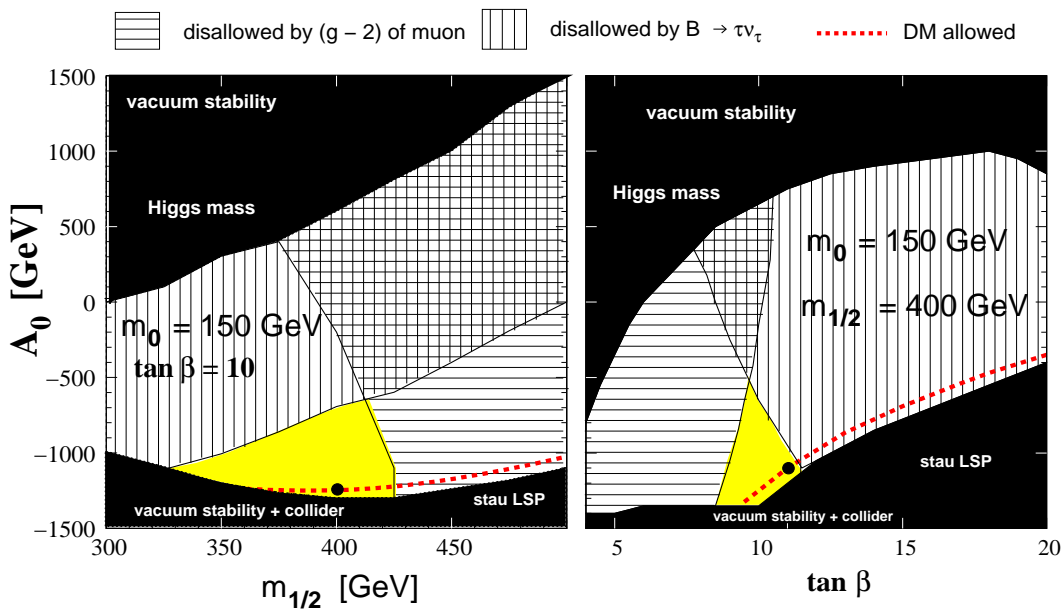} }
\caption{{\small Further constraints on the cMSSM parameter space. The 
left panel shows constraints on the $m_{1/2}$--$A_0$ plane, for $m_0 = 
150$~GeV and $\tan\beta = 10$. The right panel shows, similarly, 
constraints on the $\tan\beta$--$A_0$ plane for $m_0 = 150$~GeV and 
$m_{1/2}=400$~GeV. Notations and conventions are the same as in 
Figs.~\ref{mSUGRA-1} and \ref{mSUGRA-2}. The dotted (red) line 
represents the dark-matter-compatible region, and the black dot 
superposed on it is a benchmark point chosen for LHC studies. }}
\label{mSUGRA-3}
\end{figure*}

If we glance at the right panel of Fig.~\ref{mSUGRA-2}, where $\tan\beta 
= 20$, we see that there is no allowed region at all, the disallowed 
regions showing substantial overlap and covering the whole of the plot 
area. Once again, we surmise that high values of $\tan\beta$ are 
disfavored, whatever value of $A_0$ is chosen, and that the allowed 
region in the cMSSM parameter space must lie in the neighborhood of 
$\tan\beta = 10$. We have already mentioned that consistency with the 
$B^+ \to \tau^+ \nu_\tau$ constraint requires large negative values of 
the $A_0$ parameter, which can drive $M_+$ to higher values even for the 
low $\tan\beta$ ($\approx 10)$. In order 
to see this, we plot, in the left panel of Fig.~\ref{mSUGRA-3}, the same 
constraints in the plane of $A_0$ and $m_{1/2}$, keeping $m_0$ fixed at 
the value $m_0 = 150$~GeV, for $\tan\beta = 10$. This particular value 
of $m_0$ has been chosen since in the left panel of Fig.~\ref{mSUGRA-2}, 
it lies more-or-less near the center of the allowed triangle 
(yellow/light gray shading) and is roughly the value for which the 
maximum range of $m_{1/2}$ appears to be allowed. The dimensions of this 
allowed triangle also encourage us, in Fig.~\ref{mSUGRA-3}, to ``zoom in'' 
on the range $m_{1/2} = 300$--$500$~GeV, outside which we get disallowed 
regions. However, $A_0$ is varied between $-1.5$~TeV and $+1.5$~TeV, to 
adequately cover the whole range allowed by the firm constraints, as is 
apparent from the left panel of Fig.~\ref{mSUGRA-3}.

It is immediately apparent from the left panel of Fig.~\ref{mSUGRA-3} 
that about half of the region with positive values of $A_0$ is ruled out 
by the firm constraints, and the remaining half by the $B^+ \to \tau^+ 
\nu_\tau$ measurement. The latter has a severe impact on the $A_0 < 0$ 
region as well, essentially forcing us to consider large negative values 
of $A_0$ for small values of $m_{1/2}$. Including the muon $(g-2)/2$ 
constraint, which disfavors large values of $m_{1/2}$, then clinches the 
issue, permitting only another small wedge-shaped (yellow/light gray) 
region allowed by all the constraints. The maximum range of $m_{1/2}$ 
permitted by all the constraints is around 325--425~GeV, which matches 
tolerably well with the vertical limits of the allowed triangle in 
Fig.~\ref{mSUGRA-2}, as should be the case. The value $A_0$, on the 
other hand, is quite strictly restricted to the approximate range 
$-625$~GeV to $-1.4$~TeV.
   
Of course, the above results are only for a fixed value $\tan\beta = 
10$. Though we have already seen that jumping to a much larger value 
$\tan\beta = 20$ does not lead to any allowed region, it is interesting 
to `zoom in' to the $\tan\beta$--$A_0$ plane and see the impact of all 
these constraints there. This is shown in the right panel of 
Fig.~\ref{mSUGRA-3}, where we set $m_0 = 150$~GeV as before, and 
$m_{1/2} = 400$~GeV, which lies close to its value on the left panel for 
which the allowed range of $A_0$ is maximum. Once again, the combined 
constraints predict a small allowed region, with a maximum range of 
$\tan\beta$ lying roughly between 8 and 12. If we now refer to 
Fig.~\ref{B_tau_nu}($b$), this means that the charged Higgs boson is 
predicted to have a mass $M_+ > 600$~GeV.

Combining all these results, therefore, we obtain a roughly polyhedral 
allowed region in the four-dimensional parameter space, which is 
enclosed in a rather small hypercube with sides approximately at
\begin{eqnarray}
100~{\rm GeV} \lesssim & m_0 & \lesssim 225~{\rm GeV} \; , \nonumber \\
375~{\rm GeV} \lesssim & m_{1/2} & \lesssim 425~{\rm GeV} \; ,\nonumber \\
-1.4~{\rm TeV} \lesssim & A_0 & \lesssim -625~{\rm GeV} \; , \nonumber \\
8 \lesssim & \tan\beta & \lesssim 12 \; .
\label{allowed_box}
\end{eqnarray}
The volume of the actual allowed region is considerably smaller than 
that of the hypercube, given that the two-dimensional projections shown 
in the previous two figures are roughly triangular in shape. Compared to 
the large regions considered in traditional work on the cMSSM, this 
constitutes a rather specific region of parameter space, and encourages 
us to make specific predictions based on this model. One can easily 
argue that the qualitative features of the mass spectrum and couplings 
will not undergo dramatic changes from one end to the other of so small 
a box as this one, unless, indeed, it encloses some point(s) of 
instability. This is unlikely, for none of the many studies of the cMSSM 
parameter space have ever shown such a possibility.

The very first prediction one would naturally demand from a specific 
point or region in the cMSSM parameter space is whether this can 
adequately explain the dark-matter content of the Universe as a relic 
density of LSP's. The CMBR data indicate a relic density $\Omega h^2 = 
0.1123 \pm 0.007$ at 95\% C.L. \cite{Komatsu:2010fb}.  In general, SUSY 
models with a low-lying mass spectrum, such as the one in question, tend 
to predict too large a density of LSP's unless these are coannihilated 
by some reaction with a substantial crosssection. This leads to a 
restriction on the cMSSM parameter space, which, given the accuracy of 
the CMBR data, confines us, more or less, to a line passing through the 
four-dimensional parameter space. The dark-matter requirement is known 
to favor large negative values of $A_0$ \cite{Bednyakov:1996jr, 
*Bednyakov:1996ax, *Bottino:2000jx, *Bednyakov:2002dz, *Ellis:2003ry, 
*Stark:2005mp, *Calibbi:2007bk, *Chattopadhyay:2007di}, and it is rather 
gratifying to see that this line passes right through the allowed region 
in the parameter space discovered in this work -- which seems to 
indicate that a SUSY explanation of dark-matter may indeed be the 
correct one. The line consistent with the dark-matter 
requirement\footnote{We do not go so far as to call it a constraint, 
though this is not unheard of in the literature.} is shown by red dots 
on both panels in Fig.~\ref{mSUGRA-3} and may be seen to pass clearly 
through the allowed region, favoring a narrow range of $A_0$ around 
$-1.25$~TeV and $\tan\beta$ in the range $9 - 11$. This region is rather 
close to the forbidden stau LSP region, as is apparent in both panels of 
Fig.~\ref{mSUGRA-3}. In the allowed region, the lighter stau 
$\widetilde{\tau}_1$ is rendered very light due to the presence of a 
large negative $m_\tau A_\tau$ in the off-diagonal terms in the stau 
mixing matrix; however it is marginally heavier than the neutralino 
$\widetilde{\chi}_1^0$ LSP. This permits stau coannihilation with 
neutralinos, and reduces the relic density so that it is within the 
observed range.

We note that there does not seem to be any \'a priori reason for the 
region allowed by the low-energy constraints to match with the dark 
matter-compatible region, since the low-energy constraints come from 
processes quite different from those that control the relic density. 
Nevertheless, the fact that the two regions do show some overlap  
encourages us to argue that we are now converging on the correct region 
in the hitherto-unknown parameter space. We may, therefore, make bold as 
to venture some predictions regarding the collider signals for this 
range of parameter space, especially in the context of the LHC.

In order to make a clear prediction about the LHC signals, we choose the 
following benchmark point in the cMSSM parameter space:
\be
m_0 =  150~{\rm GeV} \; , \quad
m_{1/2}  = 400~{\rm GeV} \; , \quad
A_0  =  - 1250~{\rm GeV} \; , \quad
\tan\beta  =  10 \; , \quad
\mu  >  0 \; .
\label{benchmark-cMSSM}
\ee
Not only does this lie inside the hypercube marked out in 
Eq.~(\ref{allowed_box}), but it lies well within the allowed region, and 
right on the line corresponding to the dark-matter requirement. In 
Fig.~\ref{mSUGRA-3}, this benchmark point is indicated by a small black 
circle in both panels [on top of the dotted (red) line labeled ``dark 
matter'']. The mass spectrum and signals expected for this ``golden point'' 
will be typical of the entire allowed region, which is, after all, 
rather small.
At this benchmark point, we get the central values of the 
observables to be 
$\mathrm{BR}(B \to X_s \gamma) = 2.64 \times 10^{-4},
R^{NP}_{\tau \nu}=0.993$, and $a_\mu = 13.0 \times 10^{-10}$.
Clearly, all of these are consistent with the measurements to
within $2\sigma$, though $R^{NP}_{\tau\nu}$ only barely
survives the $2\sigma$ bound.

Let us first discuss the cMSSM mass spectrum expected with this 
benchmark point. We calculate\footnote{The masses obtained using 
different RG evolution algorithms differ by a few GeV, and the errors 
from the calculation are difficult to quantify. Here we give the exact 
values obtained by {\sc SUSY-HIT}.} the mass spectrum and the branching 
ratios using the code {\sc SUSY-HIT} \cite{susyhit} and taking $m_t = 
173.1$ GeV. The lightest Higgs boson $h^0$ is predicted to have a mass 
around 119~GeV, which is consistent with current bounds, but lies 
precisely in the range where its detection is most problematic because 
of large QCD backgrounds at the LHC. In fact, a light Higgs boson of 
this mass range must be detected through the rare decay $h^0 \to 
\gamma\gamma$, which is unlikely in the 7~TeV run, and will require the 
accumulation of a fair amount of statistics even in the 14~TeV run. The 
heavy Higgs bosons, including the $H^+$, will lie in the range 
835--840~GeV, which is again kinematically inaccessible in the 7~TeV 
run, but may be detectable at 14~TeV. We have already shown that for 
$\tan\beta = 10$, as taken for this benchmark point, this high value of 
$M_+$ is what allows us to evade the $B^+ \to \tau^+ \nu_\tau$ 
constraint. Turning now to sparticles, the LSP $\widetilde{\chi}_1^0$ 
will have a mass of 164~GeV, with the next-to-LSP (NLSP) being, as 
expected, the $\widetilde{\tau}_1$ with a mass of 171~GeV. As explained 
above, the closeness in these masses permits the coannihilation of 
stau, so that the relic density is controlled. In this scenario, this 
stau and the lightest neutralino are the only sparticles with masses 
below that of the top quark, all other particles being heavier. The 
nearly degenerate lighter chargino $\widetilde{\chi}_1^+$ and second 
neutralino $\widetilde{\chi}_2^0$ lie at 315~GeV, while the other 
sleptons and the sneutrinos have different masses in the 200--320~GeV 
range. The gluino mass, however, is as high as 934~GeV and the squark 
masses mostly populate the range 800--900~GeV, except for the 
$\widetilde{b}_1$, with mass around 719~GeV, and a light stop 
$\widetilde{t}_1$ which lies as low as 393~GeV.

An immediate consequence of these large squark and gluino masses is that 
the sparticle production cross section at the LHC will be on the low 
side: at 7~TeV it will be around 0.4~pb at the leading order (LO), while 
at 14~TeV, it will have the much healthier value of 5.2~pb at LO. About 
60\% of these crosssections come from squark pair production, of which 
roughly half is due to $\widetilde{t}_1 \widetilde{t}_1^\ast$ production 
alone. The $\widetilde{t}_1$ will decay to a top quark and a neutralino 
with a BR $\sim 2/3$, and hence, a possible signal would be a 
top-enriched final state with large missing transverse energy (MET). 
However, the enormous $t\bar{t}$ background to this process must be 
taken into consideration when studying this signal. The other 
traditional signals for SUSY -- cascade decays of the gluino or squarks 
to charginos and heavy neutralinos, ending up in multileptons, jets, and 
MET -- in this case provide $\tau$-rich final states because of the 
low-lying $\widetilde{\tau}_1$. However, $\tau$'s coming from the decay 
$\widetilde{\tau}_1 \to \tau + \widetilde{\chi}_1^0$ will generally be 
too soft for detection, because of the small splitting 
$M(\widetilde{\tau}_1) - M(\widetilde{\chi}_1^0) \simeq 7$~GeV. Final 
states involving other charged leptons will be suppressed. This 
indicates that the best option to seek SUSY with this benchmark point is 
the final state with four or more jets and substantial MET, which can 
arise from cascade decays involving only strongly interacting sparticles 
and the invisible LSP.

\begin{figure*}[ht]
\centerline{\epsfxsize= 7.0 cm \epsfysize= 7.0 cm \epsfbox{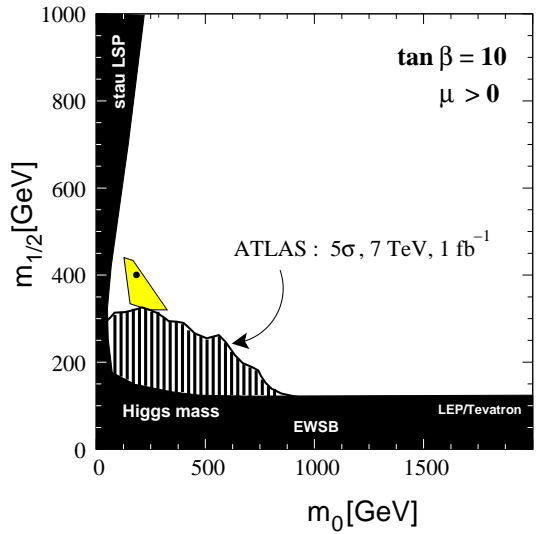} }
\caption{{\small The $m_{0}$--$m_{1/2}$ plane in the cMSSM, showing the 
allowed region (yellow/light gray) for floating $A_0$, as well as the 
$5\sigma$ reach of the ATLAS detector at the 7~TeV run, using the jets + 
MET signal. The black dot inside the allowed region is the golden point 
discussed in the text. The entire nonshaded region inside this plot is 
accessible to the LHC with $\sqrt{s} = 14$~TeV with a luminosity of 
10~fb$^{-1}$.
}}
\label{fig:collider_reach}
\end{figure*}

In Fig.~\ref{fig:collider_reach} we show the allowed parameter space 
(yellow/light gray) in the $m_0$--$m_{1/2}$ plane for $\tan\beta = 10$ 
and floating $A_0$ (as in Fig.~\ref{mSUGRA-2}), and also the ATLAS 
$5\sigma$ discovery limit \cite{ATLAS} at the 7~TeV run with an 
integrated luminosity of 1~fb$^{-1}$, using the four-jets + MET channel. 
It may be seen that the entire parameter space allowed by low-energy 
constraints at 95\% C.L., including our golden point, lies just outside 
the $5\sigma$ discovery limit of ATLAS. The ATLAS study, in fact, has 
shown that at neighboring points, an overall crosssection of about 
1~pb is required for a $5\sigma$ discovery using the four-jets + MET 
channel. Making a simpleminded scaling with the predicted crosssection 
of 0.4~pb at our benchmark point, {\it one may expect a signal in this 
channel at the level of about $2\sigma$}. The same conclusions can be 
reached using the CMS 95\% exclusion plot for the 7~TeV run 
\cite{CMSnote}. Thus, by the end of 2011, as per the LHC running 
schedule, we may begin to see tantalizing hints of SUSY. If this should 
occur, then, in the 14~TeV run, it will be easy to see a $5\sigma$ 
signal with even 1~fb$^{-1}$ of data -- which should collect within the 
first few months. Interestingly, some of the direct production modes for 
charginos, which are electroweak in nature, lie at the level of 5\%--10\% 
of the total crosssection. These may be difficult to detect in the 
7~TeV run, but in the 14~TeV run, they are sure to provide additional signals 
for SUSY. With such copious production of sparticles, the LHC could 
indeed act as a SUSY factory, as mentioned in the Introduction.


\section{NUHM : explaining $\mathbf{B^+ \to \tau^+ \nu_\tau}$}

In the previous analysis, we have seen that the combination of 
constraints on the cMSSM parameter space leads to the prediction of a 
small value of $\tan\beta$ and hence, according to 
Fig.~\ref{B_tau_nu}($b$), the charged Higgs boson is necessarily heavy. 
Comparison with Fig.~\ref{B_tau_nu}($a$) readily shows that in this 
limit, the model is only just consistent with the $B^+ \to \tau^+ 
\nu_\tau$ constraint at 95\% C.L. However, if we take the position that 
the $2\sigma$ discrepancy between the SM and the experimental result 
should be {\it explained} by a positive NP contribution, then the cMSSM 
fails the test, for it actually tends to diminish the SM prediction, and 
barely survives exclusion in a decoupling limit. This bare survival, by 
the skin of its teeth, as it were, is the proximate cause of the 
stringent constraints on the cMSSM parameter space discussed in the 
previous section.
 
As the cMSSM is the SUSY model with the maximum number of simplifying 
assumptions (and hence the minimum number of free parameters), it is 
interesting to ask if the relaxation of one or more of these assumptions 
could lead to a SUSY model which actually explains, rather than merely 
remains consistent with, the $B^+ \to \tau^+ \nu_\tau$ discrepancy. 
Since the NP effect in $B^+ \to \tau^+ \nu_\tau$ involves the scalar 
sector of the cMSSM, an obvious option would be to consider a model 
where the parameters of the Higgs sector are given a greater degree of 
flexibility than in the highly constrained cMSSM. In this context, an 
obvious choice of model is the so-called nonuniversal Higgs-mass (NUHM) 
model, which is an extension of the cMSSM where the Higgs-mass 
parameters $m_{H_1}$ and $m_{H_2}$ are delinked from the universal 
scalar mass parameter $m_0$ at the GUT scale and are allowed to vary 
freely \cite{Berezinsky:1995cj}. At the electroweak scale, these two 
extra parameters $m_{H_1}$ and $m_{H_2}$ are usually traded for the 
Higgsino mixing parameter $\mu$ and the pseudoscalar Higgs boson mass 
$m_A$. This model, therefore, has {\it six} parameters, viz. $m_0, 
m_{1/2}, \mu, M_A, A_0$ and $\tan\beta$.

NUHM models have been studied rather extensively, and various constraints 
on the six-dimensional parameter space have been found and exhibited in 
the literature \cite{Eriksson:2008cx, Ellis:2002wv, *Ellis:2002iu, 
*Baer:2005bu, *Ellis:2009ai, Buchmueller:2009fn}. What interests us here 
is the fact that $M_A$ is a free parameter in the model, and it can be 
easily exchanged for $M_+$, to which it is related by the well-known 
SUSY relation
\begin{equation}
M_+^2 = M_A^2 + M_W^2 \; ,
\label{eq:sum-rule}
\end{equation}
at tree level. We can accordingly fix $m_0$, $m_{1/2}$, $A_0$, etc. at 
whatever value is required to satisfy the other constraints in the 
cMSSM, and then claim an explanation for the $B^+ \to \tau^+ \nu_\tau$ 
discrepancy by choosing a low $M_+$ and a high $\tan\beta$ -- this 
freedom being allowed by the bigger parameter space in the theory. 
However, large values of $\tan\beta$ and small values of $M_+$ lead to 
large charged Higgs boson-mediated contributions to the FCNC process 
$B_s \to \mu^+\mu^-$, thus restricting the freedom in choosing parameter 
values. Here, as explained earlier, SUSY cancellations between the 
charged Higgs boson-mediated and the gaugino-mediated contributions come 
to the rescue: stringent bounds can be evaded if the gaugino masses are 
somewhat low, comparable to that of the light charged Higgs boson $H^+$. 
This, in turn, demands that the universal gaugino mass $m_{1/2}$ be 
somewhat small, compared with the other parameters, which are not so 
restricted.

\begin{figure*}[ht]
\centerline{ \epsfxsize= 12.0 cm \epsfysize= 7.0 cm \epsfbox{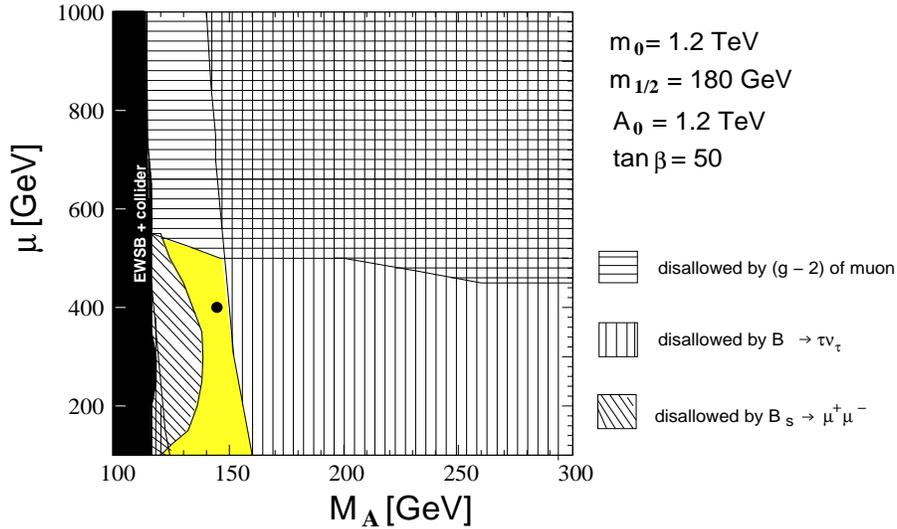} }
\caption{{\small Constraints on the $M_A$--$\mu$ parameter space in the 
NUHM model. Notations and conventions are the same as in the previous 
plots, except that now there is a significant constraint from $B_s \to 
\mu^+\mu^-$ rather than $B_d \to X_s \gamma$. The yellow/light gray 
region is allowed by all the constraints, and the black dot inside it is 
a benchmark point chosen for LHC studies. }}
\label{NUHM}
\end{figure*}

While an exhaustive study of the NUHM parameter space vis-\'a-vis the 
present set of constraints would require a separate work in itself, it 
is interesting to see if the NUHM model can at all provide regions in 
parameter space which are consistent with all the constraints, and can 
simultaneously provide a NP explanation of the $B^+ \to \tau^+ \nu_\tau$ 
discrepancy.  To illustrate that this is, in fact, possible, we show in 
Fig.~\ref{NUHM} the regions allowed by the different constraints in the 
$m_A$--$\mu$ plane, keeping all the other parameters fixed at
\begin{equation}
m_0 = 1.2~{\rm TeV} \; , \qquad m_{1/2} = 180~{\rm GeV} \; ,
\qquad A_0 = 1.2~{\rm TeV} \; ,
\qquad \tan\beta = 50 \; . 
\label{fixedpar}
\end{equation}
A glance at the figure will reveal that here, as in the cMSSM, there is 
a complementarity between the $B^+ \to \tau^+ \nu_\tau$ constraint and 
the $(g - 2)/2$ constraint, the former tending to rule out larger values 
of $M_A$ and the latter tending to rule out larger values of $\mu$, as a 
result of which only a small rectangular patch in the $m_A$--$\mu$ plane 
is allowed by both constraints taken together. A large portion of this 
remaining patch is again disallowed by the $B_s \to \mu^+\mu^-$ 
constraint, leaving a roughly sickle-shaped yellow/light gray region. In 
this region, $M_A$ remains in the approximate range 100--150~GeV; i.e. 
$M_+$ lies roughly in the range 125--170~GeV, according to 
Eq.~(\ref{eq:sum-rule}).  Figure~\ref{B_tau_nu} then tells us that this is 
not only consistent with the experimental data, but it is precisely the 
range for which the NP explanation saturates the gap between the SM 
prediction and the experimental central value.

As before, to be precise about the LHC signals, we choose a benchmark 
point, which has the fixed parameter choices of Eq.~(\ref{fixedpar}) as 
well as
\be
M_A =  145~{\rm GeV} \; , \quad\qquad \mu = 400~{\rm GeV} \; , 
\label{benchmark-NUHM}
\ee
which is indicated in Fig.~\ref{NUHM} by a small black dot in the 
middle of the allowed (yellow/light gray) patch. 
The central values of the observables at this point are
$\mathrm{BR}(B \to X_s \gamma) = 3.50 \times 10^{-4},
R^{NP}_{\tau\nu} = 1.24, 
\mathrm{BR}(B_s \to \mu^+ \mu^-) = 3.22 \times 10^{-8},
a_\mu = 12.8 \times 10^{-10}$, all of which are
well within the $2\sigma$ range of the respective measurements.
We note that the relic 
density of LSP's at this point is not enough to saturate the CMBR 
requirements, which means that this model is not ruled out by the 
latter, but is not a solution to that problem either.

The major features of the mass spectrum at this benchmark point are as 
follows: the lightest Higgs boson lies just beyond the LEP disallowed 
region, at 112~GeV, and, as in the cMSSM, this is a difficult mass range 
to search for the lightest Higgs boson. We would have to wait for enough 
statistics to accumulate at the 14~TeV run to see this Higgs boson in 
the $\gamma\gamma$ channel. The $H^0$ and $A^0$ lie at 145~GeV and may 
just be detectable through their decays to $WW^\ast$ modes, while the 
charged Higgs boson $H^+$ lies at 170~GeV, where it will decay to 
$\tau^+\nu_\tau$. These may also be detectable fairly early in the 
14~TeV run. The LSP, as before, is the lightest neutralino 
$\widetilde{\chi}_1$ with a mass of 71~GeV, which is permitted by the 
LEP direct search bound as applied to the NUHM \cite{Amsler:2008zzb}. 
The $\widetilde{\chi}_1^+$ and $\widetilde{\chi}_2^0$ lie at around 
130~GeV, while the other gauginos are heavier than 400~GeV. The sleptons 
and squarks in this model are very heavy, lying in the range 700~GeV to 
1.2~TeV, but the gluino $\widetilde{g}$ is comparatively light, having a 
mass of 511~GeV.

As a consequence of the low-lying gaugino states contrasted with heavy 
sfermions, the dominant sparticle production channels in this model turn 
out to be to chargino pairs $\widetilde{\chi}_1^+\widetilde{\chi}_1^-$ 
($\sim 50\%$) and chargino-neutralino pairs 
$\widetilde{\chi}_1^\pm\widetilde{\chi}_2^0$ ($\sim 25\%$), with gluino 
pairs $\widetilde{g}\widetilde{g}$ bringing up the rear ($\sim 20\%$). 
The total crosssection in this model would be 4.4~pb at 7~TeV and 
27.3~pb at 14~TeV, i.e. much larger than the earlier case of the cMSSM. 
The gluino production channel can give rise to the same jets + MET 
signal as before, as each gluino will undergo three-body decays through 
virtual squarks. The production crosssection for 
$\widetilde{g}\widetilde{g}$ pairs at 7~TeV is around 0.8~pb, which 
indicates that the jets + MET signal may actually be observable in the 
7~TeV run at the 3$\sigma$--4$\sigma$ level when 1~fb$^{-1}$ of data have been 
collected.  At 14~TeV, of course, a few hundred pb$^{-1}$ of data would 
be enough to obtain a $5\sigma$ signal in this channel. Turning now to 
the chargino production modes, the rate of production of 
$\widetilde{\chi}_1^+\widetilde{\chi}_1^-$ indicates a crosssection for 
a dilepton + MET signal around 90~fb, which may not be discernible above 
the background, especially as the mass splitting 
$M(\widetilde{\chi}_1^+) - M(\widetilde{\chi}_1^0)$ is rather small. 
However, the $\widetilde{\chi}_1^\pm\widetilde{\chi}_2^0$ channel could 
lead to hadronically quiet trilepton + MET signals at the level of 
74~fb, which have smaller SM backgrounds and hence could probably be 
seen as more data are collected in the 7~TeV run, and would be a 
sure-shot option at the 14~TeV run.

Before concluding this section, we reiterate that NUHM models which 
explain the $B^+ \to \tau^+ \nu_\tau$ discrepancy and, at the same time, 
remain consistent with the data on $B_s \to \mu^+\mu^-$, will 
generically come with light gauginos, and lead to collider signals 
somewhat similar to those discussed above. However, what we have studied 
is just one portion of the NUHM parameter space, inasmuch as we fixed 
$m_0$ and $A_0$ to very large values. A more comprehensive scan over the 
NUHM parameter space might reveal more patches consistent with all the 
constraints, and some of these may lead to collider signals which are 
different from those discussed in the context of our benchmark point. 
The detailed exploration of the NUHM parameter space in this context 
calls for a separate study.


\section{Concluding remarks}

With the commissioning of the LHC, the search for new physics beyond the 
Standard Model has assumed paramount importance in particle physics at 
the high scale. However, low-energy observables from flavor physics, 
like those from the decays of $K$, $D$, or $B$ mesons, can offer indirect 
constraints on high scale physics. Indeed, with the high statistics 
available at the $B$ factories BaBar and Belle, the freedom available 
for new physics has been substantially constrained. Most of the low-
energy measurements have been consistent with the SM, and hence allow 
only a little leeway for NP. On the other hand, it is seen that the 
handful of measurements that indicate a $\sim 2 \sigma$ deviation from 
the SM also restrain the NP parameters from taking arbitrary values.
 
In this paper, we have shown that the combined effect of both these 
kinds of low-energy measurements -- those consistent with the SM (e.g. 
the branching ratios of $B_d \to X_s \gamma$ and $B_s \to \mu^+ \mu^-$) 
as well as those showing deviations from the SM (e.g. the anomalous magnetic 
moment of the muon, and the branching ratio of the $B^+ \to \tau^+ \nu_\tau$) -- 
results not only in indicating new physics, but also in pinpointing the 
relevant new physics parameters. In particular, we have pointed out that 
the latest measurement of $B^+ \to \tau^+ \nu_\tau$ branching ratio has 
a large impact on a large class of NP models, especially those which 
include a charged Higgs boson $H^+$. In fact, the decay $B^+ \to \tau^+ 
\nu_\tau$, by itself, can constrain most of the models with minimal 
flavor violation that involve an $H^+$. This is because the latest 
measurement gives a branching ratio $\sim 2 \sigma$ more than the SM 
prediction. If this discrepancy is to be explained by a MFV model, one 
needs very light charged Higgs bosons ($M_+ \lesssim 200$ GeV) and large 
$\tan\beta$ ($\gtrsim 20$). On the other hand, a heavy charged Higgs 
boson ($M_+ \gtrsim 300$ GeV) and a small $\tan\beta$ can be barely 
consistent with the data to within $2 \sigma$, but cannot be considered 
an explanation for the gap between theory and experiment. This is a 
general result that can be applied to any member of the MFV models, and 
we choose to apply it to the constrained MSSM, which is motivated by 
mSUGRA and is one of the most predictive SUSY models.

In cMSSM models, the charged Higgs boson is typically heavy, so that 
only the low $\tan\beta$ region survives the $B^+ \to \tau^+ \nu_\tau$ 
measurement. When combined with the anomalous magnetic moment of the 
muon, the fate of even this region is in jeopardy: indeed, for a 
vanishing universal trilinear coupling $A_0$, there is no region in the 
cMSSM parameter space that is consistent with both these measurements to 
95\% C.L.. The situation can only be salvaged with a large and negative 
$A_0$, and that too for an extremely small region in the 
$m_0$--$m_{1/2}$ plane. The combined low-energy data thus pinpoint us to 
a very specific location (the golden point) in the five-dimensional 
parameter space of cMSSM: $\mu >0, ~A_0 \approx -1.25 \textrm{~TeV}, 
~\tan\beta \approx 10 \; , m_0 \approx 150 \textrm{~GeV}, m_{1/2} 
\approx 400 \textrm{~GeV}$, with a spread not more than that given in 
Eq.~(\ref{allowed_box}). It is remarkable that for part of this 
specific region, including the golden point, the mass and coupling of 
the LSP are exactly such that it can account for all the dark-matter in 
the Universe. This may either be a coincidence, or an indication that we 
are on the right track in our quest.
 
If we indeed are on the right track, and the golden point of the cMSSM 
is actually the NP that we have all been looking for, then we may not 
have to wait too long for its discovery. Since the values of $m_0$ and 
$m_{1/2}$ at this point are rather small, at the LHC, one expects a weak 
2$\sigma$ signal in the jets + MET channel even in the 7~TeV run with 
1~fb$^{-1}$ of integrated luminosity, and a $5 \sigma$ discovery early 
in the 14 TeV run with just 1~fb$^{-1}$ of data.

While the above suggestive coincidence is quite appealing, and the 
prospects of the detection of SUSY during the early parts of the 14 TeV 
run quite enticing, even at the golden point the model barely survives 
the 95\% limit bounds and does not offer any help at all in explaining 
the $B^+ \to \tau^+ \nu_\tau$ data -- the absence of a light $H^+$ makes 
it impossible for the cMSSM to do so. We therefore explore a related but 
less constrained model, the NUHM, where the charged Higgs boson mass can 
be considered to be a free parameter. Here the presence of an extra 
parameter works wonders for explaining the low-energy data, covering the 
entire experimentally allowed region, including the central value. This 
model can also lead to rather spectacular trilepton + MET signals at the 
LHC, which may become detectable even towards the end of the 7~TeV run.

While our work tends to indicate a rather specific region of parameter 
space and specific signals at the LHC, especially for the cMSSM, there 
are some caveats which need to be taken into consideration, even 
apart from theoretical issues in the construction of the cMSSM. The 
first is the issue of experimental errors on the low-energy 
measurements, which we have taken at the 2$\sigma$ level. If these are 
given more latitude (e.g. taken at the 3$\sigma$ level) the constraints 
from low-energy processes would be considerably relaxed. In particular, 
the $B^+ \to \tau^+ \nu_\tau$ measurement would still allow wide regions 
in the cMSSM parameter space. However, it would still disfavor very 
large values of $\tan\beta \sim 50$. A more serious point is the 
asymptotic behavior $R^{\rm NP}_{\tau \nu_\tau} \to 1$ in the large 
$M_+$ limit, as compared to the 2$\sigma$ bound $R_{\tau 
\nu_\tau}^{\mathrm{NP}} > 0.99$. The strong constraint on NP comes 
because one must squeeze the contribution of the charged Higgs bosons 
into the narrow region $0.99-1.00$. A small downward revision in the 
lower bound on $R^{\rm NP}_{\tau \nu_\tau}$ could allow large 
$\tan\beta$ values even for large $M_+$. Such deviations can come from a 
variety of sources, such as higher-order corrections, a slightly-changed 
value of $f_B$ or $|V_{ub}|$, or a revised experimental result. On the 
other hand, a small upward revision of the allowed $R^{\rm NP}_{\tau 
\nu_\tau}$ band could rule out the entire gamut of MFV models with $M_+ 
> 200$~GeV. In particular, even the small leeway allowed for the cMSSM 
would then be closed. We note, therefore, that the bounds and predictions 
presented in this paper are specific to the experimental limits as they 
stand at present.

We have not made a very detailed study of the LHC signals, confining 
ourselves to generalities, because it is somewhat premature, at this 
stage, to make very definite predictions in this regard. Nevertheless, 
our work has highlighted the fact that if indeed we are to accept the 
cMSSM at face value, as most LHC studies do, then we should take the 
cMSSM in its entirety, i.e. all constraints from all sectors, 
including the low-energy sector. The next year and the years after it 
will be the most crucial in determining if our analysis, in fact, is on 
the right track.
  
\section*{Acknowledgements}

We would like to thank G.~Mohanty (Belle Collaboration) for initiating a 
discussion on $B ^+ \to \tau^+ \nu_\tau$ at WHEPP-XI. 


%
\end{document}